\documentclass{aip-cp}

\usepackage[numbers,sort&compress]{natbib}
\usepackage{rotating}
\usepackage{graphicx}
\usepackage{amssymb}

\newcommand{\pmb}[1]{\mathbf{#1}}

\begin{document}

\title{Role of the Symmetry Energy on the Structure of Neutron Stars with Unified Equations of State}
\author[aff1]{Nicolas Chamel\corref{cor1}}
\author[aff2]{John Michael Pearson}
\corresp[cor1]{Corresponding author and speaker: nchamel@ulb.ac.be}
\author[aff3]{Alexander Y. Potekhin}
\author[aff4,aff1]{Anthea F. Fantina}
\author[aff5]{Camille Ducoin}
\author[aff6]{Anup Dutta}
\author[aff1]{St\'ephane Goriely}
\author[aff1]{Lo\"ic Perot}
\affil[aff1]{Institut d'Astronomie et d'Astrophysique, CP-226, Universit\'e
Libre de Bruxelles, 1050 Brussels, Belgium}
\affil[aff2]{D\'ept. de Physique, Universit\'e de Montr\'eal, Montr\'eal
(Qu\'ebec), H3C 3J7 Canada}
\affil[aff3]{Ioffe Institute, Politekhnicheskaya 26,
194021 St. Petersburg, Russia}
\affil[aff4]{Grand Acc\'el\'erateur National d'Ions Lourds (GANIL), CEA/DRF -
 CNRS/IN2P3, Boulevard Henri Becquerel, 14076 Caen, France}
\affil[aff5]{Institut de Physique Nucl\'eaire de Lyon, CNRS/IN2P3, Universit\'e
 Claude Bernard Lyon 1, Villeurbanne, France}
\affil[aff6]{School of Physics, Devi Ahilya University, Indore, 452001 India}
\maketitle

\begin{abstract}
The role of the symmetry energy on the internal constitution and the global structure 
of a cold nonaccreted neutron star is studied using a set of unified equations of state. 
Based on the nuclear energy-density functional theory, these equations of state provide 
a thermodynamically consistent treatment of all regions of the star and were calculated 
using the four different Brussels-Montreal functionals BSk22, BSk24, BSk25 and BSk26. 
Our predictions are compared to various constraints inferred from astrophysical observations 
including the recent detection of the gravitational wave signal GW170817 from a binary 
neutron-star merger. 
\end{abstract}

\section{INTRODUCTION}
\label{sec1} 

Formed in the aftermath of gravitational core-collapse supernova explosions, neutron stars are initially fully fluid and very hot with internal temperatures as high as $10^{10}-10^{11}$~K~\cite{hae07}. In such a furnace, the compressed stellar material is likely to undergo all kinds of nuclear and electroweak processes eventually reaching full thermodynamical equilibrium. The proto-neutron stars cool rapidly by emitting neutrinos. After a few months, the surface of the star - possibly surrounded by a very thin atmospheric plasma layer of light elements - still remains hot enough to be liquid. However, the layers 
beneath crystallize thus forming a solid crust. After about a hundred thousand years, heat from the stellar interior diffuses to the surface and is dissipated in the form of radiation.  Assuming that full thermodynamical equilibrium is maintained throughout the different cooling stages, the interior of an old  neutron star consists of ``cold catalyzed matter'', i.e., electrically charge-neutral matter in its absolute ground state with density exceeding that found inside the heaviest atomic nuclei~\cite{hw58,htww65}. 
This hypothesis can reasonably be expected to be valid in any neutron star that has not accreted material 
from a stellar companion, but will otherwise fail because of the relative slowness with which accreted matter reaches nuclear equilibrium. Although properties of hot nuclear matter off equilibrium can be explored in terrestrial laboratories up to a few times saturation density $n_0\simeq 0.16$ nucleons~fm$^{-3}$ by analysing heavy-ion collisions, the extreme conditions prevailing in the very dense and highly neutron-rich interior of a cold neutron star cannot be experimentally reproduced.

Neutron stars thus offer the unique opportunity to probe novel phases of matter~\cite{bc18}. In particular, the interior of a neutron star can be decomposed into three distinct regions below its thin 
atmosphere and liquid ``ocean'': an outer crust made of a body-centered cubic lattice of exotic nuclei in a charge neutralizing background of highly degenerate electrons, an inner crust consisting of neutron-proton clusters immersed in a neutron sea (possibly enriched with protons at sufficiently high densities), and a liquid core of nucleons and leptons. Other particles such as hyperons might be present in the most massive neutron stars~\cite{hae07,bc18}, but we do not consider this possibility here. Because the core of a neutron star contains asymmetric nuclear matter at densities up to $\sim10 n_0$, astrophysical observations may shed light on the longstanding issue of the density dependence of the symmetry energy, defined here as the difference between the energy per nucleon of pure neutron matter (NeuM) and that of symmetric nuclear matter (SNM). To this end, we have developed a series of four unified equations of state (EoS) in the framework of the nuclear energy-density functional (EDF) 
 theory~\cite{pea18}. These EoS provide a thermodynamically consistent description of all regions of a neutron star. The underlying EDFs were precision fitted to essentially all experimental atomic mass data and were simultaneously adjusted to realistic NeuM EoS as obtained from many-body calculations. They mainly differ in their predictions for the symmetry energy. These EoS are used to study the role of the symmetry energy on the structure of a cold nonaccreted neutron star. 

\section{NUCLEAR ENERGY-DENSITY FUNCTIONAL THEORY}
\label{sec2} 

\subsection{Formalism}

The nuclear energy-density functional (EDF) theory aims at providing a universal description of various nuclear systems, from atomic nuclei to extreme astrophysical environments 
such as neutron stars and supernova cores (see, e.g. Ref.~\cite{duguet14} for a review). 

In this theory, the energy $E$ of a nuclear system (assumed here to be invariant under time reversal) is expressed as a universal functional of the so-called normal and abnormal density matrices defined by~\cite{doba84}
\begin{equation}
n_q(\pmb{r}, \sigma; \pmb{r^\prime}, \sigma^\prime) = <\Psi|c_q(\pmb{r^\prime},\sigma^\prime)^\dagger c_q(\pmb{r},\sigma)|\Psi>\, ,
\end{equation}
\begin{equation}
\widetilde{n_q}(\pmb{r}, \sigma; \pmb{r^\prime}, \sigma^\prime) = -\sigma^\prime <\Psi|c_q(\pmb{r^\prime},-\sigma^\prime) c_q(\pmb{r},\sigma)|\Psi>\, , 
\end{equation}
respectively where $|\Psi>$ is the many-body wave function, $c_q(\pmb{r},\sigma)^\dagger$ and $c_q(\pmb{r},\sigma)$ are the creation and destruction operators 
for nucleons of charge type $q$ ($q=n,p$ for neutron, proton respectively)  at position $\pmb{r}$ with spin coordinate $\sigma=\pm1$. The ground-state is obtained by minimizing the energy $E$ under the constraint 
of fixed numbers of neutrons and protons. The minimization is usually carried out by expressing the density matrices in terms of auxiliary independent two-component quasi-particle wavefunctions $\psi^{(q)}_{1k}(\pmb{r}, \sigma)$ and $\psi^{(q)}_{2k}(\pmb{r}, \sigma)$, as  
\begin{equation}
 n_q(\pmb{r}, \sigma; \pmb{r^\prime}, \sigma^\prime) =
\sum_{k(q)}\psi^{(q)}_{2k}(\pmb{r}, \sigma)\psi^{(q)}_{2k}(\pmb{r^\prime}, \sigma^\prime)^* 
\end{equation}
and
\begin{equation}
\widetilde{n_q}(\pmb{r}, \sigma; \pmb{r^\prime}, \sigma^\prime) =
-\sum_{k(q)}\psi^{(q)}_{2k}(\pmb{r}, \sigma)
\psi^{(q)}_{1k}(\pmb{r^\prime}, \sigma^\prime)^*=-
\sum_{k(q)}\psi^{(q)}_{1k}(\pmb{r}, \sigma)\psi^{(q)}_{2k}(\pmb{r^\prime},
\sigma^\prime)^* \, ,
\end{equation}
where the index $k$ represents the set of suitable quantum numbers, the symbol $*$ denotes complex conjugation, and the quasiparticle states are subject to the following conditions~\cite{doba84}: 
\begin{equation}
 \sum_{\sigma} \int d^3r\, \Biggl\{  n_q(\pmb{r_1}, \sigma_1; \pmb{r}, \sigma)
 \widetilde{n_q}(\pmb{r}, \sigma; \pmb{r_2}, \sigma_2)
-  \widetilde{n_q}(\pmb{r_1}, \sigma_1; \pmb{r}, \sigma)
n_q(\pmb{r}, \sigma; \pmb{r_2}, \sigma_2) \Biggr\} 
= 0\, , 
\end{equation}
\begin{equation}
 \sum_{\sigma} \int d^3r\, \Biggl\{  n_q(\pmb{r_1}, \sigma_1; \pmb{r}, \sigma)
 n_q(\pmb{r}, \sigma; \pmb{r_2}, \sigma_2)
 +  \widetilde{n_q}(\pmb{r_1}, \sigma_1; \pmb{r}, \sigma)\widetilde{n_q}(\pmb{r}, \sigma; \pmb{r_2}, \sigma_2)\Biggr\} 
  = n_q(\pmb{r_1}, \sigma_1; \pmb{r_2}, \sigma_2)\, .
\end{equation}

Although the EDF theory can potentially describe the exact ground state of the system, the corresponding EDF is unknown. Phenomenological functionals have been traditionally constructed  
from density-dependent effective nucleon-nucleon interactions in the framework of the self-consistent mean-field methods~\cite{bhr03}. However, it should be stressed that the resulting EDFs actually include beyond mean-field effects (with respect to ``bare'' nucleon-nucleon interactions) through the fitting to nuclear data. Although such a formulation imposes stringent restrictions on the form of the EDF, it guarantees the cancellation of self-interaction errors~\cite{cha10} (nonetheless, the EDFs may still be contaminated by \emph{many-body} self-interactions errors, see, e.g. Ref.~\cite{duguet14} and references therein). The Brussels-Montreal EDFs considered here are based on generalized Skyrme effective interactions of the form~\cite{cgp09}
\begin{eqnarray}\label{eq:skyrme}
v(\pmb{r}_{i},\pmb{r}_{j}) & = & t_0(1+x_0 P_\sigma)\delta({\pmb{r}_{ij}})+\frac{1}{2} t_1(1+x_1 P_\sigma)\frac{1}{\hbar^2}\left[p_{ij}^2\,
\delta({\pmb{r}_{ij}}) +\delta({\pmb{r}_{ij}})\, p_{ij}^2 \right]\nonumber\\
&+&t_2(1+x_2 P_\sigma)\frac{1}{\hbar^2}\pmb{p}_{ij}\cdot\delta(\pmb{r}_{ij})\,\pmb{p}_{ij}+\frac{1}{6}t_3(1+x_3 P_\sigma)n(\pmb{r})^\alpha\,\delta(\pmb{r}_{ij})
\nonumber\\
&+& \frac{1}{2}\,t_4(1+x_4 P_\sigma)\frac{1}{\hbar^2} \left[p_{ij}^2\,
n({\pmb{r}})^\beta\,\delta({\pmb{r}}_{ij}) +
\delta({\pmb{r}}_{ij})\,n({\pmb{r}})^\beta\, p_{ij}^2 \right] \nonumber\\
&+&t_5(1+x_5 P_\sigma)\frac{1}{\hbar^2}{\pmb{p}}_{ij}\cdot n({\pmb{r}})^\gamma\,\delta({\pmb{r}}_{ij})\, {\pmb{p}}_{ij}  \nonumber\\
& +&\frac{\rm i}{\hbar^2}W_0(\pmb{\hat\sigma_i}+\pmb{\hat\sigma_j})\cdot
\pmb{p}_{ij}\times\delta(\pmb{r}_{ij})\,\pmb{p}_{ij} \, , 
\end{eqnarray}
where $\pmb{r}_{ij} = \pmb{r}_i - \pmb{r}_j$, $\pmb{r} = (\pmb{r}_i + 
\pmb{r}_j)/2$, $\pmb{p}_{ij} = - {\rm i}\hbar(\pmb{\nabla}_i-\pmb{\nabla}_j)/2$
is the relative momentum, $\pmb{\hat\sigma_i}$ and $\pmb{\hat\sigma_j}$ are Pauli spin matrices, 
 $P_\sigma$ is the two-body spin-exchange operator, and $n(\pmb{r})$ denotes the average nucleon number 
density. The $t_4$ and $t_5$ terms were originally introduced to remove spurious spin and spin-isospin instabilities~\cite{cg10}. Nuclear pairing is treated using a different effective interaction of the form
\begin{equation}\label{eq:pairing}
 v(\pmb{r}_{i},\pmb{r}_{j})=\frac{1}{2}(1- P_\sigma)f^{\pm}_q v^{\pi\, q}[n_n(\pmb{r}),n_p(\pmb{r})]\delta(\pmb{r}_{ij}) \, ,
\end{equation}
 where $n_n(\pmb{r})$ and $n_p(\pmb{r})$ denote the average neutron and proton number densities respectively. The pairing strength  $v^{\pi\, q}[n_n(\pmb{r}),n_p(\pmb{r})]$ given by~\cite{cha10b}
 \begin{equation}
v^{\pi\,q}[n_n,n_p]=-\frac{4\pi^2\hbar^2}{k_{{\rm F}q}M_q}\biggl[ 2\log\left(\frac{2\varepsilon^{(q)}_{\rm F}}{\Delta_q}\right)+ \Lambda\left(\frac{\varepsilon_\Lambda}{\varepsilon^{(q)}_{\rm F}}\right) \biggr]^{-1}\, ,
\end{equation}
\begin{equation}
\Lambda(x)=\log (16 x) + 2\sqrt{1+x}-2\log\left(1+\sqrt{1+x}\right)-4\, ,
\end{equation}
where $M_q$ is the nucleon mass, $k_{{\rm F}q}= (3\pi^2 n_q)^{1/3}$ is the Fermi wave number, $\varepsilon^{(q)}_{{\rm F}} = \hbar^2 k_{{\rm F}q}^2/(2 M_q)$, and $\varepsilon_\Lambda$ is a cutoff in the single-particle energy above the Fermi level,  
was determined in such a way as to reproduce the $^1S_0$ pairing gaps $\Delta_q$ of NeuM and SNM as obtained from extended Brueckner-Hartree-Fock calculations including medium-polarization effects~\cite{cao06} (see Refs.~\cite{cgp08,gcp09,gcp09b} for details). But because of Coulomb effects and a possible charge-symmetry breaking of nuclear forces, the proton pairing strength may be slightly different from the
 neutron pairing strength. This fine-tuning was made possible by introducing the parameters $f^{\pm}_q$. Our neglect of the time-odd fields implicit in our use of the equal-filling approximation was compensated phenomenologically by allowing the pairing force to also depend on whether there is an even ($+$) or odd ($-$)  number of nucleons of the charge type in question. By definition, $f_n^+=1$.  

With such kind of interactions, the energy $E$ can be expressed as 
 \begin{equation}
  E=\int d^3r\, \mathcal{E}(\pmb{r})\, ,
 \end{equation}
where the local energy density $\mathcal{E}(\pmb{r})$ depends on the local normal and abnormal nucleon number densities, respectively, 
\begin{equation}\label{eq:rhoq}
n_q(\pmb{r}) = \sum_{\sigma=\pm 1}n_q(\pmb{r}, \sigma; \pmb{r}, \sigma)\, ,
\end{equation}
\begin{equation}\label{eq:rhoqtilde}
\widetilde{n_q}(\pmb{r}) = \sum_{\sigma=\pm 1}\widetilde{n_q}(\pmb{r}, \sigma; \pmb{r}, \sigma)\, ,
\end{equation}
the kinetic-energy density 
\begin{equation}\label{eq:tauq}
\tau_q(\pmb{r}) = \sum_{\sigma=\pm 1}\int\,{\rm d}^3\pmb{r^\prime}\,\delta(\pmb{r}-\pmb{r^\prime}) \pmb{\nabla}\cdot\pmb{\nabla^\prime}
n_q(\pmb{r}, \sigma; \pmb{r^\prime}, \sigma)\, , 
\end{equation}
and the spin-current vector density
\begin{eqnarray}\label{eq:Jq}
\pmb{J}_q(\pmb{r}) &=& -{\rm i}\sum_{\sigma,\sigma^\prime=\pm1}\int\,{\rm d}^3\pmb{r^\prime}\,\delta(\pmb{r}-\pmb{r^\prime}) 
\pmb{\nabla} n_q(\pmb{r}, \sigma; \pmb{r^\prime},
\sigma^\prime) \times  \pmb{\hat\sigma}_{\sigma^\prime\sigma}   \nonumber \\
&=&{\rm i}\sum_{\sigma,\sigma^\prime=\pm1}\int\,{\rm d}^3\pmb{r^\prime}\,\delta(\pmb{r}-\pmb{r^\prime}) 
\pmb{\nabla^\prime} n_q(\pmb{r}, \sigma; \pmb{r^\prime},
\sigma^\prime) \times  \pmb{\hat\sigma}_{\sigma^\prime\sigma}\, .
\end{eqnarray}
Note that all the terms in $J^2$ and $J_q^2$ from the energy density are dropped as discussed in Ref.~\cite{gcp10}. 
Minimizing the energy leads to the self-consistent Hartree-Fock-Bogoliubov (HFB) equations~\cite{doba84}
\begin{eqnarray}
\label{eq:HFB}
\sum_{\sigma^\prime=\pm1} h_q(\pmb{r} )_{\sigma \sigma^\prime}\psi^{(q)}_{1k}(\pmb{r},\sigma^\prime) + \Delta_q(\pmb{r}) \psi^{(q)}_{2k}(\pmb{r},\sigma)=(E_k+\mu_q) \psi^{(q)}_{1k}(\pmb{r},\sigma)\, , \nonumber \\
 \Delta_q(\pmb{r})\psi^{(q)}_{1k}(\pmb{r},\sigma)  -\sum_{\sigma^\prime=\pm1} h_q(\pmb{r})_{\sigma \sigma^\prime} \psi^{(q)}_{2k}(\pmb{r},\sigma^\prime) =(E_k-\mu_q)\psi^{(q)}_{2k}(\pmb{r},\sigma) \, , 
\end{eqnarray}
where $E_k$ denotes the quasiparticle energy, $\mu_q$ the nucleon chemical potential, 
\begin{equation}
h_q(\pmb{r})_{\sigma\sigma^\prime} = -\pmb{\nabla}\cdot \frac{\delta E}{\delta\tau_q(\pmb{r})}\pmb{\nabla}
 \delta_{\sigma\sigma^\prime}+ \frac{\delta E}
{\delta n_q(\pmb{r})} \delta_{\sigma\sigma^\prime}
-{\rm i}\frac{\delta  E}
{\delta \pmb{J_q}(\pmb{r})} \cdot \pmb{\nabla} \times \pmb{\hat\sigma}_{\sigma\sigma^\prime} 
\end{equation}
represents the single-particle Hamiltonian,  
\begin{equation}
\Delta_q(\pmb{r})=\frac{\delta E}{\delta \widetilde{n_q}(\pmb{r})}=\frac{1}{2}v^{\pi q} [n_n(\pmb{r}),n_p(\pmb{r})]\widetilde{n_q}(\pmb{r}) 
\end{equation}
is a potential responsible for pairing. Expressions for these fields can be found in Refs.~\cite{cgp09,cha10b}. 
Depending on the choice of boundary conditions, Equation~(\ref{eq:HFB}) can describe the atomic nuclei present in the outer crust of a neutron star, the inhomogeneous nuclear matter constituting the inner crust or the homogeneous neutron-proton mixture in the core. However, a reliable description of all these 
different nuclear phases requires a suitable adjustment of the parameters of the EDF. 

\subsection{Brussels-Montreal energy-density functionals}

The parameters of the Brussels-Montreal EDF series BSk22-BSk26~\cite{gcp13} were determined primarily by fitting to 
the 2353 measured masses of atomic nuclei having $Z,N\geq8$ from the 2012 Atomic Mass Evaluation~\cite{ame12}. 
Theoretical nuclear masses were calculated within the HFB method allowing for axial deformations~\cite{sam02}. 
In an attempt to account for dynamical correlations, the following estimate for the spurious 
collective energy was subtracted from the HFB energy (see Ref.~\cite{cgp08} for a discussion)
\begin{equation}
\label{eq:collective}
E_\mathrm{coll}= E_\mathrm{rot}^\mathrm{crank}\Big\{b~\tanh(c|\beta_2|) + d|\beta_2|~\exp\{-l(|\beta_2| - \beta_2^0)^2\}\Big\} \,  ,
\end{equation}
in which $E_\mathrm{rot}^\mathrm{crank}$ denotes the cranking-model value of the rotational correction and $\beta_2$ the quadrupole deformation, 
while all other parameters were fitted freely. This correction was shown to be in good agreement with calculations using 5D collective Hamiltonian~\cite{gcp10}.
To the HFB energy, a phenomenological Wigner correction was also added  
\begin{equation}\label{eq:wigner}
E_W = V_W\exp\Bigg\{-\lambda\,\Bigg(\frac{N-Z}{A}\Bigg)^2\Bigg\}
+V_W^{\prime}|N-Z|\exp\Bigg\{-\Bigg(\frac{A}{A_0}\Bigg)^2\Bigg\} \quad .
\end{equation}
This term contributes significantly only for light nuclei ($A < A_0$) or nuclei with $N$ close to $Z$ (see, e.g., Ref.~\cite{cgp08} for a discussion 
of the physical interpretation of these terms). The spurious centre-of-mass energy was removed following an essentially exact procedure~\cite{sg03}. 
Moreover, a correction for the finite size of the proton was made to both the 
charge radius and the energy~\cite{cgp08}. Finally, Coulomb exchange for protons was 
dropped, thus simulating neglected effects such as Coulomb correlations, charge-symmetry breaking, and vacuum polarization~\cite{gp08}. 

\begin{figure}
  \centerline{\includegraphics[width=.9\textwidth]{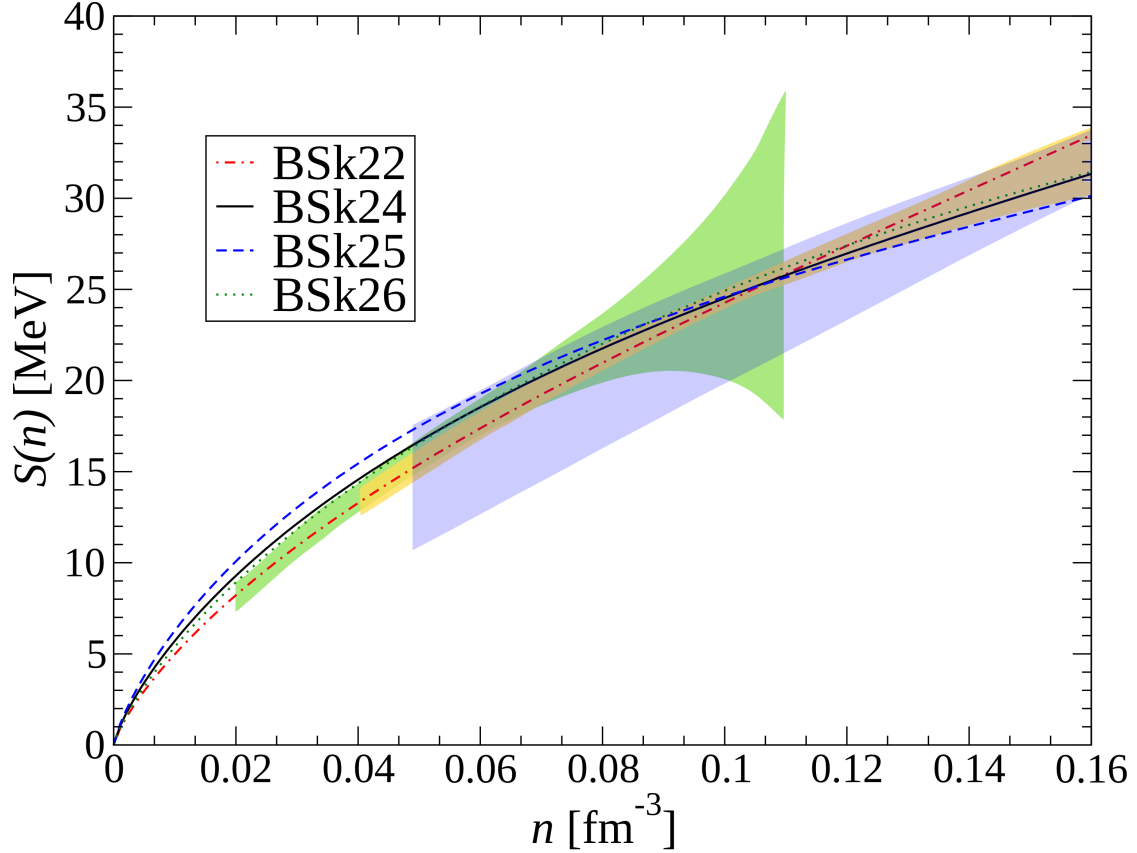}}
\caption{Variation of the symmetry energy $S(n)$ with density $n$ for the Brussels-Montreal functionals. The shaded areas are experimental constraints: from heavy-ion collisions~\cite{tsang09} (blue), from isobaric-analog states and neutron skins~\cite{dan14} (yellow), and from the electric dipole polarizability of $^{208}$Pb~\cite{zha15} (green).}
  \label{esym}
\end{figure}

\begin{figure}
  \centerline{\includegraphics[width=.9\textwidth]{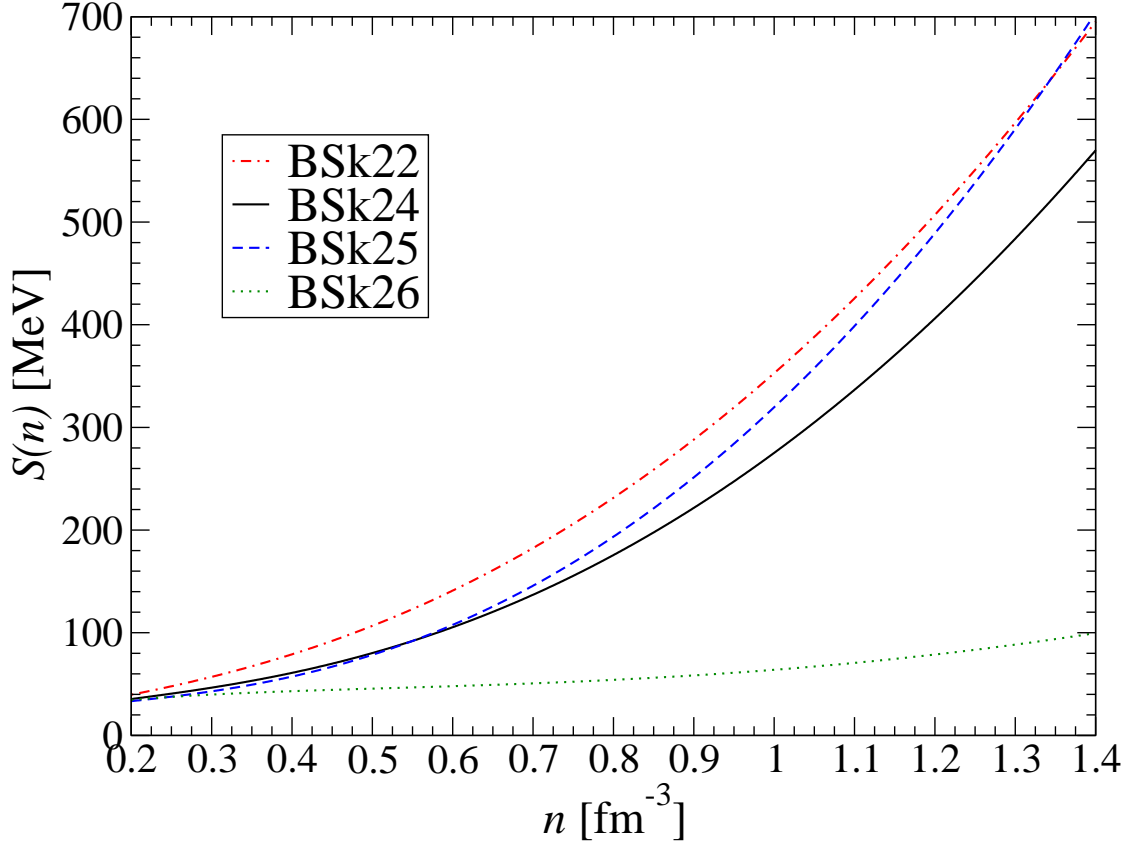}}
\caption{Variation of the symmetry energy $S(n)$ at densities $n>n_0$ for the Brussels-Montreal functionals.}
  \label{esym2}
\end{figure}

To ensure reliable extrapolations to the highly neutron-rich and very dense interiors of neutron stars,  the Brussels-Montreal EDFs were further constrained to reproduce the EoS of homogeneous NeuM, as calculated by many-body theory. Although the EoS is fairly well determined at subsaturation densities, it remains highly uncertain at supersaturation densities prevailing in the core of the most massive neutron stars. Two different EoS were considered: the rather stiff EoS labelled as `V18' by Li and Schulze \cite{ls08} 
and the softer EoS labelled as `A18 + $\delta\,v$ + UIX$^*$' by Akmal, Pandharipande, and Ravenhall \cite{apr98}.
The fit to nuclear masses along with these constraints do not lead to a unique determination of the EDF. 
Expanding the energy per nucleon of infinite nuclear matter (INM) of density $n = n_0(1 + \epsilon)$ and charge asymmetry $\eta = (n_n - n_p)/n$ about the equilibrium density $n=n_0$ and $\eta = 0$, 
\begin{equation}
e(n, \eta) = a_v + \left(J + \frac{1}{3}L\epsilon\right)\eta^2 +
\frac{1}{18}(K_v + \eta^2K_\mathrm{sym})\epsilon^2 + \cdots 
\end{equation} 
the incompressibility coefficient $K_v$ was further restricted to lie in the experimental range $K_v=240\pm10$~MeV~\cite{col04}. 
To achieve a good fit to nuclear masses with a root-mean-square deviation as low as $0.5-0.6$ MeV, it was necessary to limit the allowed values of the symmetry-energy coefficient $J$ from 29 to 32 MeV. The EDFs BSk22, BSk23, BSk24 and BSk25 were thus all fitted to the NeuM EoS of Ref.~\cite{ls08} while having $J$ = 32, 31, 30 and 29 MeV, respectively. To assess the role of the NeuM EoS, the EDF BSk26 was fitted to the softer EoS of Ref.~\cite{apr98} with $J=30$~MeV. Nuclear-matter properties for the EDFs are summarized in Table~\ref{tab1}. The intermediate EDF BSk23 will not be further considered here. As shown in Fig.~\ref{esym}, the variation of the symmetry energy $S(n)$ with density $n$ as predicted by the Brussels-Montreal EDFs are compatible with experimental constraints from transport-model analyses of midperipheral heavy-ion collisions of Sn isotopes~\cite{tsang09}, from the analyses of isobaric-analog states and neutron-skin data~\cite{dan14}, and from the electric dipole polarizability of $^{208}$Pb~\cite{zha15}. The symmetry energy is defined here as the difference between the energy per nucleon in NeuM and the energy per nucleon in SNM,  
 \begin{equation}
 S(n) = e_\mathrm{NeuM}(n) - e_\mathrm{SNM}(n)\, .
\end{equation} 
The EDFs mainly differ in their predictions for the symmetry energy at densities $n>n_0$, as shown in Fig.~\ref{esym2}. 

\begin{table}[ht]
\caption{Nuclear-matter properties for the Brussels-Montreal functionals.}
\label{tab1}
\begin{tabular}{|c|ccccc|}\hline
&BSk22&BSk23&BSk24&BSk25&BSk26\\
\hline
$J$ {[MeV]} &32.0&31.0&30.0&29.0&30.0\\
$L$ {[MeV]} &68.5&57.8 &46.4&36.9&37.5\\
$K_v$ {[MeV]} &245.9&245.7 &245.5 & 236.0 & 240.8\\
$K_\mathrm{sym}$ {[MeV]} &13.0& -11.3 & -37.6 & -28.5 & -135.6\\
\hline
\end{tabular}
\end{table}

\section{UNIFIED EQUATIONS OF STATE OF COLD CATALYZED MATTER}
\label{sec3} 

\subsection{Methods}

For densities greater than $\sim10^{-10}$ nucleons fm$^{-3}$ ($\sim10^{6}$ g cm$^{-3})$ atoms are completely ionised by the pressure and electrons form a degenerate Fermi gas. Since nuclear degeneracy likewise holds everywhere it follows that the cold catalyzed matter hypothesis is valid throughout the star except in a thin layer at densities $\rho\lesssim10^6$ g cm$^{-3}$, where the atomic ionization and electron degeneracy can be incomplete. The properties of this outermost region have been extensively discussed in Ref.~\cite{hae07}. Therefore, we calculated the outer crust only for densities $\rho \gtrsim 10^6$~g~cm$^{-3}$ mainly as described in Ref.~\cite{pgc11} except that we used complete expressions for the electron exchange and charge polarization effects (see Ref.~\cite{pea18} for further details). The equilibrium properties were determined by minimizing the Gibbs free energy per nucleon $g$ at each given pressure $P$ assuming that each crustal layer consists of a perfect body-centered cubic crystal made of a single nuclear species (binary compounds may be present, but only at the interface between pure adjacent crustal layers, see Ref.~\cite{chf2016mix}). For the masses of nuclei, we made use of 
experimental data from the 2016 Atomic Mass Evaluation~\cite{ame16}, supplemented by the very recent measurements of copper isotopes~\cite{welker2017}. For nuclei with experimentally unknown masses, we used the 
HFB-22, HFB-24, HFB-25 or HFB-26 tables~\cite{bruslib}. Above some pressure $P_{\rm drip}$ such that $g=M_n c^2$, some neutrons drip out of nuclei marking the transition to the inner crust~\cite{cfzh15}. Full HFB calculations of the inner crust properties would be computationally extremely costly due to the necessity to account for unbound states. For this reason, we adopted the fourth-order extended Thomas-Fermi method with proton shell and pairing corrections added perturbatively using the Strutinsky integral theorem~\cite{pcgd12,pcpg15}. All details can be found in Ref.~\cite{pea18}. We recall only the main features here. Nuclear clusters were supposed to be spherical. To further speed up the computations, 
the Wigner-Seitz (WS) cell was approximated by a sphere of radius $R$. Moreover, the 
nucleon density distributions in the WS cell were parameterized as the sum of a constant ``background" term and a ``cluster" term according to 
\begin{equation}
n_q(r) = n_{Bq} + n_{\Lambda q}f_q(r)  \quad ,
\end{equation}
\begin{equation}
f_q(r) = \frac{1}{1 + \exp \left\{\Big(\frac{C_q - R}
{r - R}\Big)^2 - 1\right\} \exp \Big(\frac{r-C_q}{a_q}\Big) }
\end{equation}
(if $n_{\Lambda q}$ is negative the cluster becomes a bubble), 
where $q = n, p$ for neutrons or protons respectively, while $n_{B,q}$, $n_{\Lambda,q}$, $C_q$, and 
$a_q$, are free parameters.
This profile guarantees that the first three derivatives of the density vanish at the cell surface, as required by the usual implementation of the ETF method 
(see Section II of Ref.~\cite{pcgd12}, where other relevant details will be found). At high enough densities, free protons may appear. The onset of proton emission was determined by the condition 
\begin{equation}
\frac{\hbar^2}{2M^*_p(r=0)}\left[3\pi^2n_p(r=0)\right]^{2/3}  > U_p(r=R) - U_p(r=0)\, ,
\end{equation}
where $M^*_p(r=0)$ is the effective proton mass at the center of the cell and $U_p(r)$ is the value of the proton single-particle potential at the point $r$. In the free-proton region, the proton shell and pairing corrections were dropped entirely. For convenience, we minimized the energy per nucleon at fixed  average baryon number density. As shown in Ref.~\cite{pcgd12}, density discontinuities are negligibly small in the inner crust so that minimizing the energy per nucleon or the Gibbs free energy per nucleon yields practically the same results without having recourse to a Maxwell construction. The pressure $P$ at any mean density $\bar{n}$ was calculated semi-analytically as described in Appendix B of \cite{pcgd12}. Calculations in the inner crust were performed using the same EDF as that underlying the HFB nuclear mass model used in the outer crust. Calculations in the core of a neutron star are comparatively much simpler since the energy density and the pressure are given  by analytic expressions (see Ref.~\cite{pea18}). The equilibrium composition (allowing for the presence of electrons and muons) was determined by solving the equations of beta equilibrium under the condition of charge neutrality using the same EDF as in the crust. 

\subsection{Results}

\begin{figure}
  \centerline{\includegraphics[width=.9\textwidth]{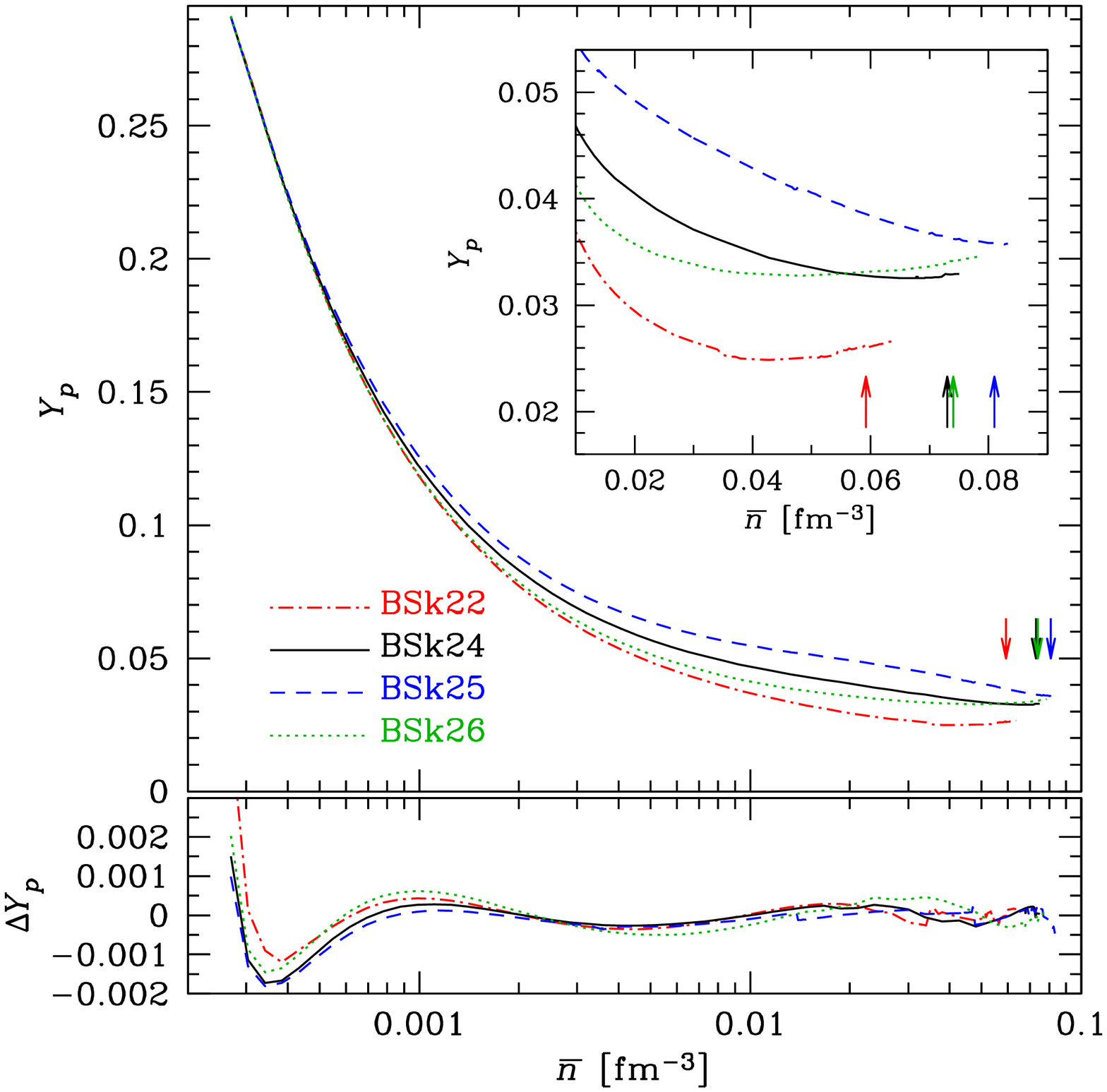}}
\caption{Upper panel: Curves show the computed equilibrium proton fraction $Y_{\rm p} = Z_{\rm eq}/A$ in the inner crust as a function of mean baryon density $\bar{n}$ for our four functionals; arrows indicate onset of proton drip. The inset shows the cross-over between BSk24 and BSk26 just before proton drip. Lower panel: Deviations between the computed data and the fitted analytic function ($\Delta Y_{\rm p}$ = fit $-$ data).}
  \label{Yic}
\end{figure}

\begin{figure}
\includegraphics[width=.9\textwidth]{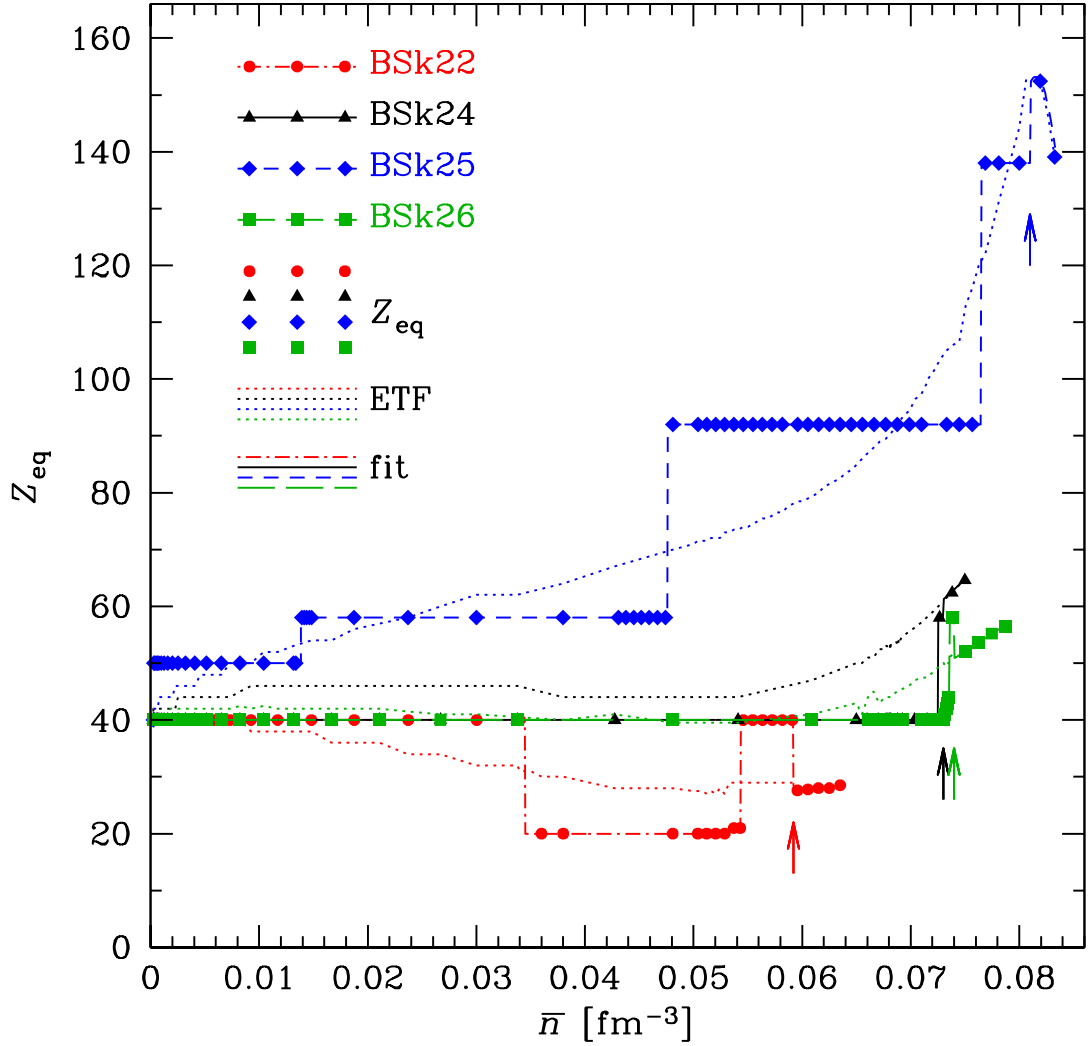}
\caption{Equilibrium value $Z_\mathrm{eq}$ of number of protons in inner crust as a function of mean baryon density $\bar{n}$ for our four functionals. For clarity only every second point is shown. Arrows indicate onset of proton drip. The dotted curves relate to the values of $Z_\mathrm{eq}$ calculated in the ETF approximation.
}
\label{Zic}
\end{figure}

\begin{figure}[ht]
 \centerline{\includegraphics[width=.9\textwidth]{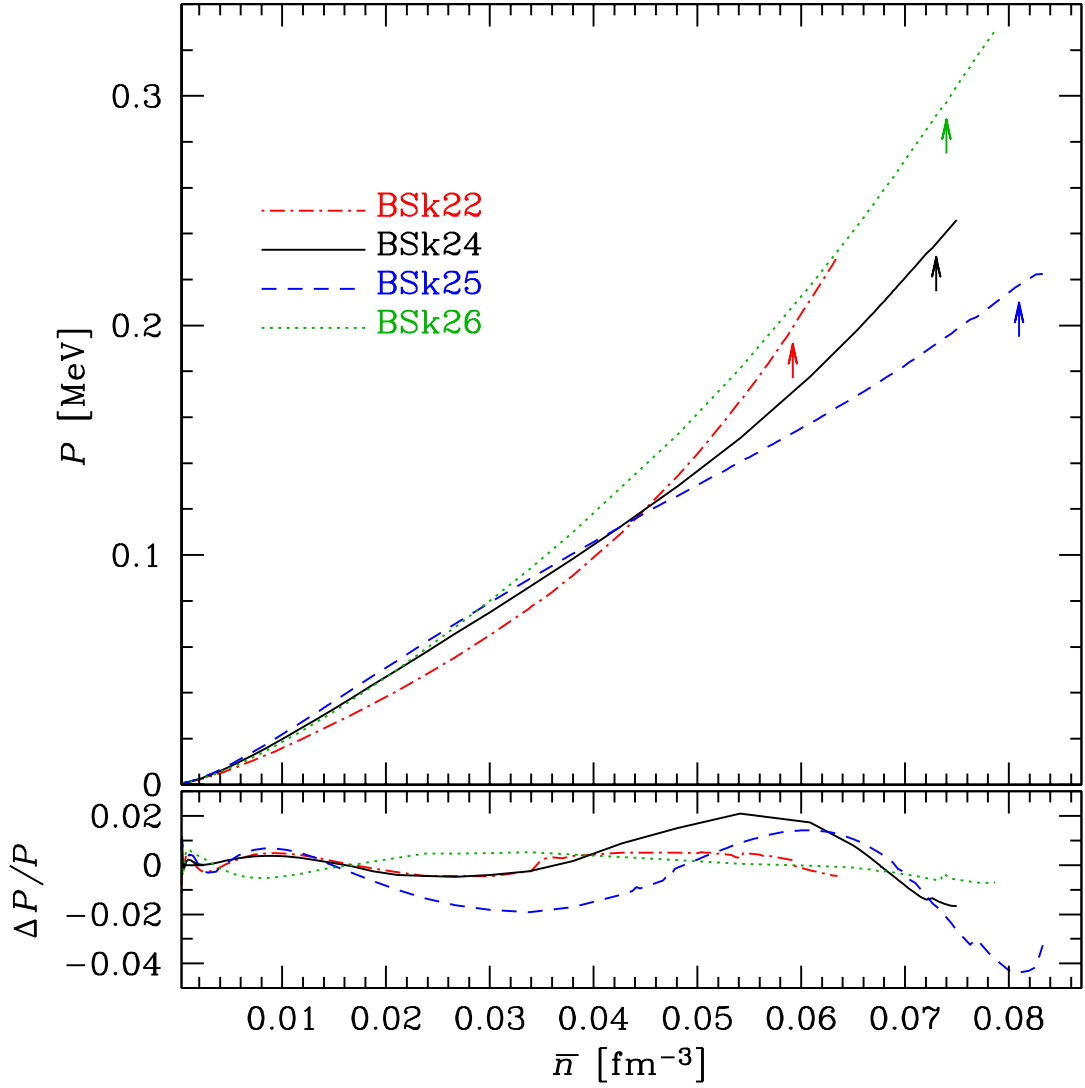}}
\caption{Upper panel: Curves show the computed pressure $P$ in the inner crust as a function of mean baryon density $\bar{n}$ for our four functionals; arrows indicate onset of proton drip. Lower panel: Fractional deviations between the computed data and the fitted analytic function
 ($\Delta P$ = fit $-$ data).}
  \label{Pic}
\end{figure}

The composition of the outer crust is completely determined by experimentally measured masses of known nuclei up to $\bar{n} \approx 3 \times 10^{-5}$~fm$^{-3}$. The use of the HFB models at higher densities introduces some uncertainties. 
Nevertheless, the differences between models remain small. Indeed, the sequence of nuclei is essentially governed by nuclear-structure effects (large regions containing nuclei with neutron magic numbers $N=50$ and $N=82$), the symmetry energy playing a minor role. The four HFB models thus lead to essentially the same variations for the proton fraction $Y_\mathrm{p} = Z/A$ and the pressure $P$ as a function of the mean density $\bar{n}$. Transitions between adjacent layers are accompanied by density discontinuities, which will be reduced (though not entirely removed) if binary compounds are allowed~\cite{chf2016mix}. In all cases, matter becomes progressively more neutron rich with increasing depth, a consequence of mechanical equilibrium and the fact that
the pressure arises mainly from the degenerate electron gas. Moreover, nuclei are less and less bound (see, e.g. Ref.~\cite{bc18}) until neutrons are emitted 
delimiting the boundary with the inner crust. The composition beyond this point is found to be more sensitive to the symmetry energy. As shown in Fig.~\ref{Yic}, the inner crust tends to be more neutron-rich with increasing $J$ reflecting the anticorrelation between $J$ and the symmetry energy $S(n)$ of INM at subnuclear densities $n < n_0$. Comparing BSk24 and BSk26, which were fitted to  the same value of $J$, shows that the choice of the constraining EoS of NeuM has no significant impact on the crustal composition. As can be seen in Fig.~\ref{Zic}, the equilibrium proton number $Z_\mathrm{eq}$ is governed to a large extent by shell effects, as in the outer crust. Apart from the well-known magic numbers 20 and 50 from atomic nuclei, the numbers 40, 58, 92, and 138 appear to be also favored in this extremely
neutron-rich environment. This change of nuclear structure mainly arises from the very strong quenching of the spin-orbit coupling. Evidence for such a quenching have been also reported in exotic nuclei \citep{sch04}. Not only does the shell structure strongly depend on the symmetry energy, but also the smooth ETF part of the energy per nucleon. In the free proton region where the shell correction is dropped entirely, $Z_\mathrm{eq}$ is thus found to be anticorrelated with $J$. Given that the proton fraction $Y_\mathrm{p}$ varies smoothly it follows that the mean number of neutrons $N_\mathrm{eq}$ per WS cell must display the same shell effects as does the proton number $Z_\mathrm{eq}$. The variation of the  pressure $P$ with $\bar{n}$ in the inner crust, shown in Fig.~\ref{Pic} for our four EDFs, directly reflects the slope of the symmetry energy. Indeed, the pressure (essentially given by free neutrons) is approximately given by   
\begin{equation}\label{Peta} 
P\approx n^2\left[\frac{K_v \epsilon}{9 n_0}  + \eta^2 \frac{dS}{dn} \right] \, ,
\end{equation}
with $\eta\sim 1$. 
Since the EDFs were all fitted to the experimental value of the incompressibility $K_v$, the differences in the pressure originate from the differences in the slope of the symmetry energy. The coincidence of the slopes of the symmetry energy at density $\sim 0.44$ fm$^{-3}$ for BSk22, BSk24 and BSk25 leads to a crossing of the pressure curves. Because the slope of the symmetry energy for BSk26 is greater than that for BSk24 at 
any density below $n_0$, the BSk26 EDF yields a stiffer EoS than BSk24 in the inner crust. For the same reason, the EoS obtained with BSk26 is softer in the core, while BSk22 leads to the stiffest EoS. Comparing only the slopes of the symmetry energy between the different EDFs, one could have expected an even softer EoS in the core for BSk26. However, the softer symmetry energy is to a large extent compensated by the higher neutron abundance such that the pressure (\ref{Peta}) in the core is not much different. A remarkable feature we found is that the
proton fraction $Y_\mathrm{p}(n)$ increases with density in the core for all EDFs, in contrast to what happens 
in both the outer and inner crusts. 

Complete numerical results and analytic fits applicable to the entire star can be found in Ref.~\cite{pea18}. 
All these fits have been implemented in Fortran subroutines,
freely available at URL http://www.ioffe.ru/astro/NSG/BSk/.

\section{GLOBAL STRUCTURE OF NEUTRON STARS}
\label{sec4} 

The global structure of a non-rotating neutron star was computed from the Tolman-Oppenheimer-Volkoff (TOV) equations~\cite{tolman1939,oppenheimer1939}
\begin{equation}
\frac{{\rm d}P(r)}{{\rm d}r} = -\frac{G\, \mathcal{E}(r)\mathcal{M}(r)}{c^2 r^2}
\biggl[1+\frac{P(r)}{\mathcal{E}(r)}\biggr] \biggl[1+\frac{4\pi P(r)r^3}{c^2\mathcal{M}(r)}\biggr]\biggl[1-\frac{2G\mathcal{M}(r)}{c^2 r}\biggr]^{-1}\, ,
\end{equation}
where $G$ is the gravitational constant, and
\begin{equation}
\mathcal{M}(r) = \frac{4\pi}{c^2}\int_0^r\mathcal{E}(r')r'^2{\rm d}r'\, .
\end{equation}
Here $\mathcal{E}(r)$ is the mass-energy density of matter at the radial coordinate $r$. The gravitational mass of the star is given by $M_\mathrm{NS}=\mathcal{M}(R_\mathrm{NS})$, where $R_\mathrm{NS}$ is the radial coordinate 
at which the radiative surface is reached.

\begin{figure} 
\includegraphics[width=.9\textwidth]{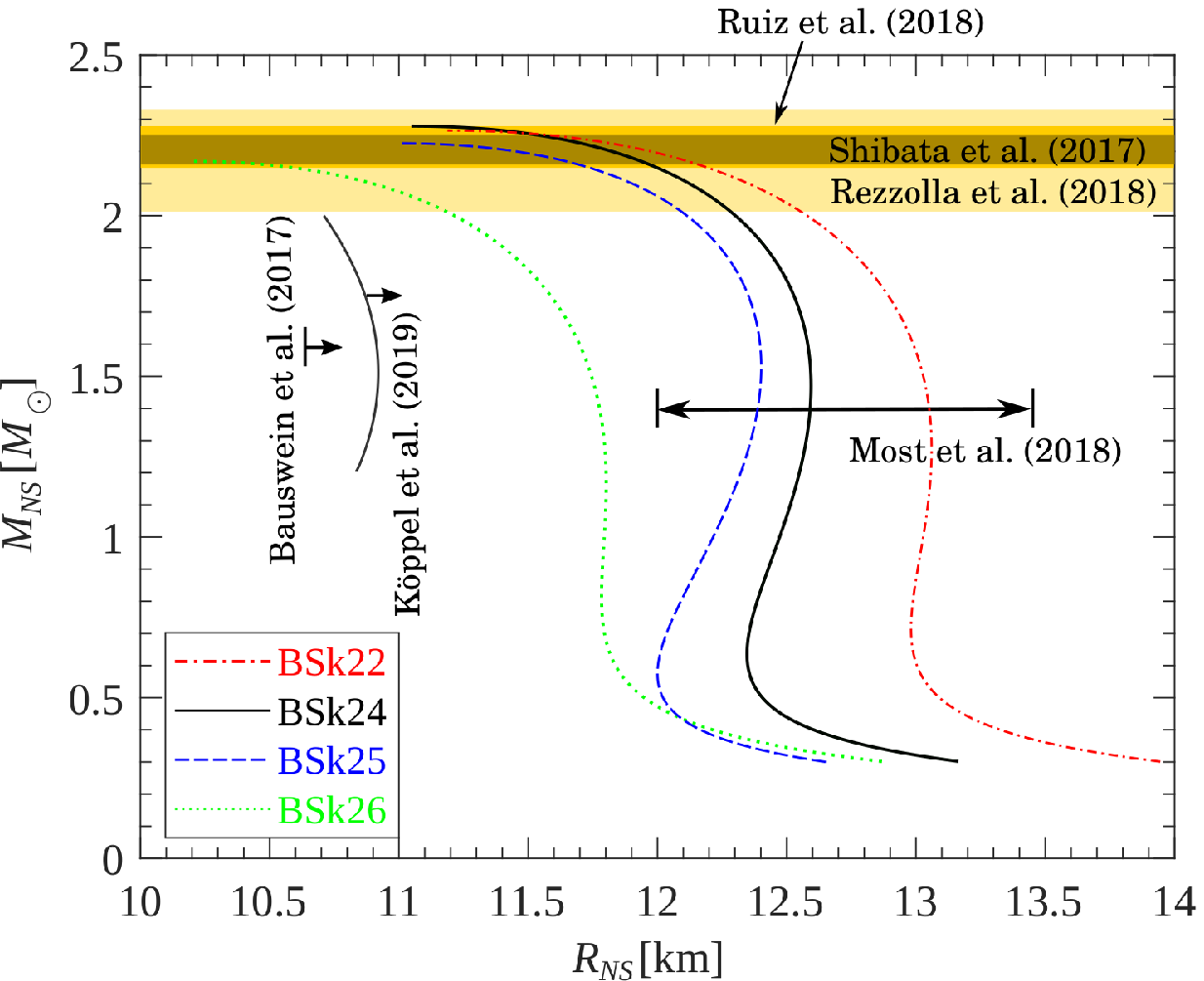}
\caption{Gravitational mass $M_\mathrm{NS}$ (in solar masses) versus circumferential radius $R_\mathrm{NS}$ of nonrotating neutron stars, calculated using the analytical EoS representations from Ref.~\cite{pea18}. Constraints inferred from the analysis of GW170817 are also shown. See text for details.} 
\label{MRBSk}
\end{figure}

The mass-radius relation for the EoSs BSk22, BSk24, BSk25, and BSk26 are shown in Figure~\ref{MRBSk}. As expected, the maximum mass is mainly determined by the NeuM constraint; BSk26 being fitted to a softer NeuM EoS than the other EDFs thus yields the lowest maximum mass. All our models are compatible with the measured masses of the most massive neutron stars known~\cite{dem10,ant13,fon16}. The figure also shows some constraints inferred from the analysis of the gravitational-wave signal GW170817 from a binary neutron-star merger. In particular, all models are compatible with the maximum-mass limits derived in Ref.~\cite{rez18} under the assumption that the binary merger product collapsed to a black hole with a mass very close to the mass-shedding limit. Quasi-universal relations between the maximum mass of nonrotating stellar models and the maximum mass supported through uniform rotation were also used. Also shown in Figure~\ref{MRBSk} are the tighter limits of Refs.~\cite{shi17,ruiz18}. All our models are consistent with the constraint of Ref.~\cite{ruiz18}, but BSk22 and BSk24 appear to be disfavored by that of Ref.~\cite{shi17}. However, these two constraints may be prone to model uncertainties as they rely on numerical simulations with further assumptions on the gamma-ray burst. 

Figure~\ref{MRBSk} shows that for a given constraining EoS of NeuM the radius is correlated with the slope $L$ of the symmetry energy, as previously found, see e.g. Refs.~\cite{fortin16,mcg18}. Lower bounds on the neutron-star radii have been obtained in Refs.~\cite{bau17,kop19} from the assumption of delayed collapse using an empirical relation for the threshold binary mass for prompt collapse. All our models satisfy these robust constraints, especially the most stringent one from Ref.~\cite{kop19}. Accounting for the observational estimate of the tidal deformability of a neutron star leads to a narrower range of allowed radii~\cite{ligo18,annala18,defin18,fatt18,most18}. The radii predicted by all our models fall below the upper limit, but BSk26 appear to agree marginally with the lower limit obtained in Ref.~\cite{most18} at 2$\sigma$ confidence level. Additional constraints can be obtained from the requirement that the extremely powerful direct Urca processes of neutrino emission operate in a small number of neutron stars (see, e.g., \cite{yakovlev04,brown18}). The direct Urca processes are allowed in sufficiently massive neutron stars. The low value of the threshold mass  $M_\mathrm{DU}\simeq 1.151 M_\odot $ given by the BSk22 EDF implies that these processes would occur in the majority of neutron stars, thus leading
to their rapid cooling, which does not agree with observations (see, e.g., Ref.~\cite{PC18}) unless these processes are suppressed by nuclear pairing in low-mass neutron stars. With threshold masses  $M_\mathrm{DU}\simeq 1.595 M_\odot $ and $M_\mathrm{DU}\simeq 1.612 M_\odot $, the BSk24 and BSk25 EDFs respectively, are consistent with neutron-star cooling observations. On the contrary, the direct Urca reactions are forbidden for all hydrostatically stable neutron stars by the BSk26 EDF, which must therefore be ruled out. 

\section{CONCLUSIONS}

We have computed a series of four different unified EoS for the cold dense matter constituting 
the interior of non accreted neutron stars within the framework of the nuclear EDF theory using the accurately calibrated Brussels-Montreal 
EDFs BSk22, BSk24, BSk25 and BSk26~\cite{gcp13}. These EDFs were precision fitted to essentially 
all the available atomic mass data (with $Z,N\geq 8$) by using the HFB method. 
In addition, these functionals were constrained to fit, up to the densities 
prevailing in neutron-star cores, two different realistic EoS of NeuM: the 
stiff EoS `V18'  of Ref.\cite{ls08} for BSk22, BSk24 and BSk25, and the softer EoS 
`A18 + $\delta\,v$ + UIX$^*$' of Ref.\cite{apr98} for BSk26. The BSk22, BSk24 and BSk25 EDFs 
mainly differ in their values of the symmetry-energy coefficient $J$, fixed to 32, 30 and 29 MeV, 
respectively. Good-quality mass fits could not be achieved outside this range, the optimum fit being obtained for $J=30$ MeV, the value also chosen for the BSk26 EDF. Although these same EDFs were used throughout the neutron star interior, we adopted different methods to implement the EDF equations in the different regions. Whereas the HFB method was used in the outer crust (except when the appropriate atomic mass data 
were available) and in the core, the computationally much faster ETFSI method was followed in the inner crust. The composition and the EoS of the inner crust and core of a neutron star are found to be very sensitive to the symmetry energy $S(\bar n)$ at densities $\bar n$ below and above normal density $n_0$ respectively. Full numerical results as well as analytic fits valid over the whole star for the all neutron-star masses are given in Ref.~\cite{pea18}. While the maximum neutron-star mass is mainly determined by the NeuM constraint, the radius is found to be correlated with the slope $L$ of the symmetry energy for a given NeuM EoS. All our EoS are compatible with the robust radius constraints of Refs.~\cite{bau17,kop19} inferred from the analysis of GW170817. However, the tighter constraints obtained from the estimated tidal deformability of neutron stars disfavor BSk26. This same model also predicts the absence of the direct Urca processes in stable neutron stars, which is difficult to reconcile with observations of some neutron stars that apparently cool via such processes.  

\section{ACKNOWLEDGMENTS}
The work of N.~C.{} was partially 
supported by the Fonds de la Recherche Scientifique - FNRS (Belgium) under 
grant n$^\circ$~CDR-J.0187.16 and CDR-J.0115.18.
J.~M.~P.{} acknowledges the partial support of the NSERC (Canada). 
The work of A.~Y.~P.{} was supported by the RFBR grant 16-29-13009-ofi-m.
 S.~G.{} acknowledges the 
support of Fonds de la Recherche Scientiffique-FNRS  (Belgium). This work was 
also partially supported by the European Cooperation in Science 
and Technology (COST) Actions MP1304 and CA16214.



\end{document}